# New technological trajectories and research directions in Cloud Computing Technology, 2004-2021

**Mario Coccia**

CNR -- National Research Council of Italy,
Collegio Carlo Alberto, Via Real Collegio, 30-10024 Moncalieri (Torino), Italy
E-mail: mario.coccia@cnr.it

**Saeed Roshani**

Allameh Tabataba'i University, Faculty of Management and Accounting,
Department of Technology and Entrepreneurship Management, Tehran, Iran
E-mail: Roshani@atu.ac.ir

**Abstract**

This goal of this study is to explore emerging trends in cloud computing technology that can support an economic and social change. We apply the methods of entity linking, which links word strings to entities from a knowledge base, to extract main keywords in cloud computing from accumulated publications from 2004 to 2021. Results suggest that in cloud computing research, "Internet of things" has an accelerated technological and scientific growth compared to the other topics. Other critical research fields that support the evolution of cloud computing are mathematical optimization and virtual machine. In particular, results suggest that computer network, encryption, big data and distributed computing are the most fast-growth fields of research in the domain of cloud computing. These findings reveal a technological competition between Cloud systems infrastructure, hardware development side, and computing and software development to play the main role in technological evolution of Cloud Computing. Moreover, this study shows that technological development of virtual machines and computing device can be of critical importance to foster an economic and technological change in many sectors. However, the implementation of cloud computing has to be supported by skill development, organizational change, and adopter engagement, to foster the management and the diffusion of cloud technologies and the exploitation of cloud-based infrastructures for competitive advantage of firms in fast-changing markets. This study can provide main information to extend the knowledge having theoretical implications to explain characteristics of the evolution of science and technology in this field of research and practical implications of innovation management for the appropriate allocation of resources towards new technological trajectories in cloud computing having a potential of growth and beneficial impact in society.

**Keywords:** Cloud Computing; Cloud Technology; Emerging Technologies; Information technology; Evolution of Technology; Topic Modeling; Entity Linking; Text analysis; Burst detection

**JEL codes:** O31, O32, O33

**Acknowledgment**

We thank Dr. Mauricio Marrone from Macquarie University for helpful suggestions and assistance. The numerical calculations were carried out on Yggdrasil at the computing cluster of the University of Geneva.

# 1. Introduction

The National Institute of Standards and Technology of the U.S. Department of Commerce (NIST, 2022) defines cloud computing as: "a model for enabling ubiquitous, convenient, on-demand network access to a shared pool of configurable computing resources (e.g., networks, servers, storage, applications and services) that can be rapidly provisioned and released with minimal management effort or service provider interaction." Cloud computing is a basic technology for managing big data and support the development of businesses (Latifian, 2022). The advantages of cloud computing technology for organizations are a fast collection of big data from different sources for decision support in markets to improve effectiveness of services. Moreover, the adoption of cloud computing technology (based on storage and information processing) has the great potential of fostering data-driven innovations in organizations with beneficial societal impact (Cresswell et al., 2022; Papazoglou and Vaquero, 2012). In fact, cloud computing technology and big data can support incremental innovations of products, processes and organizations for competitive advantage in dynamic markets with rapid changes (cf., Coccia, 2017, 2017a, 2022). Many innovations in could computing technology play a critical role for increasing productivity and quality and reducing costs for higher levels of turnover in firms after the Coronavirus Disease 2019 pandemic crisis (Ardito et al., 2021). Liu and Wang (2022) analyze China automotive companies and show that the quality of perceived information and of services in customers is due to user satisfaction in the cloud-based marketing system. Hence, cloud services in information technology are more and more a basic strategy for improving productivity and efficiency (with cost reduction), the reliability and scalability of organizational systems in markets. Ali et al. (2021) show the adoption of cloud-based service is associated with factors of compatibility and security concerns. In this context of a critical role of cloud computing for public and private firms, the goal of this study is to explore emerging trends in cloud computing technology that can explain drivers of next technological, economic and social change.

# 2. Theoretical Framework

Cloud computing is a technology having the characteristics of general-purpose technologies (GPTs) for manifold applications in various industries and organizations (Bresnahan, 2010; Coccia, 2014, Sahal, 1981). Firstly, GPTs generates new families of products/processes and support incremental and radical innovations (Helpman, 1998, p.3; Lipsey et al., 1998; cf., Coccia, 2018, 2020a). GPTs support different technological trajectories in every branch of the economy (Freeman and Soete, 1987, pp.56–57; Bresnahan and Trajtenberg, 1995, p.8; Hall and Rosenberg, 2010). Secondly, the evolution of GPTs is driven by scientific and technological change that increase the interaction between scientific fields and inter-related technological systems supporting co-evolutionary pathways (Coccia et al., 2021, 2022; Coccia and Finardi, 2013; Coccia, 2018c, 2020; Coccia and Watts, 2020; Coccia, 2018, 2018c; 2019, 2019a, 2019b; Jovanovic et al., 2021; Sun et al., 2013). GPTs, in these co-evolutionary pathways, generate a propagation of minor and major technological improvements in products and processes of many sectors, reducing costs and increasing returns-to-scale (Bresnahan and Trajtenberg, 1995; Jovanovic and Rousseau, 2005). Lipsey et al. (1998, 2005) describe other characteristics of GPTs, such as the scope for improvement, wide variety and range of uses and strong complementarities with existing or new technologies having a societal impact (cf., Coccia, 2020, 2020b, 2022a). Overall, then, GPTs are complex technologies that support product/process innovations in several sectors for a corporate, industrial, institutional, economic, and social change (Coccia, 2018a, 2020). The characteristics of GPTs can explain scientific and technological development of cloud computing research and technology in markets (Nelson, 2008, p.489; cf., Coccia, 2018a).

The goal of investigating here cloud computing technology and research is to clarify aspects of the long-run evolution of this technology and predict, whenever possible, technological trajectories having beneficial societal impact (cf.., Deshmukh and Mulay, 2021). In this context, the study here analyzes publications that are a main unit for scientific and technology analysis of cloud computing technology to show evolutionary pathways (Boyack et al., 2009). As matter of fact, bibliometric analyses of publications can capture information earlier of the cycle of technology development, unlike patent analysis (Cozzens et al., 2010; Ding et al., 2000). The idea is to analyze the evolution of cloud computing research for detecting new technological trajectories that are basic in science, technology, and society to explain the evolution of this GPT and support strategic management of R&D investments towards research fields and technologies having a high potential of growth and positive societal impact. Next section presents the methods of this scientific investigation.

## 3. Materials and Methods

### 3.1. Sample and Data Collection

To address the main question of this research, we used the Web of Science- WOS (2021) core collection database for extracting the articles related to Cloud Computing (in short, CC). The query used to extract the data from Web of Science (2021) is:

TS=("cloud comput*") OR TS=("cloud-comput*") refine By: Document Types = ("Articles or Proceeding Papers"), Language = ("English") and Publication Years = (2004-2021).

The results of this query contain 49,828 documents. This study investigates the period from 2004 to 2021 because the first year 2004 indicates the first article published in cloud computing.

### 3.2. Data Processing
- *Entity Linking*

We utilized entity linking to extract main keywords in cloud computing from accumulated publications. In this step, it is necessary to distinguish the relevant, meaningful keywords and identify the main topics that generate this research field. There are different approaches to extracting the general topics from documents using Natural Language Processing (NLP): topic modeling techniques, such as Latent Dirichlet Allocation-LDA (Blei et al., 2003), Latent Semantic Analysis-LSA (Landauer et al., 1998), probabilistic latent semantic analysis-PLSA (Hofmann, 1999), etc. These approaches have been used in recent scientific research to identify different subjects in a field of study (Coccia et al., 2021, 2022). The topic models are statistical algorithms intended to identify main themes and topics within large collections of unstructured documents (Blei, 2012). These algorithms are commonly referred to as "generative" models, as they assume that documents contain multiple topics in accordance with a statistical distribution. The documents are generated using a probabilistic document generation process that selects words from topics. The topic modelling algorithm reveals the topic of a document based on words that appear in it, their interrelationships, and how they change over time (Blei, 2012). It is possible to organize many document collections chronologically and therefore see how different topics and topic frequencies evolve over time. A topic model with time-stamped are used by researchers to capture this dynamic behavior (Chen et al., 2017). Some algorithms have been developed to mine documents chronologically (Blei and Lafferty, 2006; Wang et al., 2012; Gohr et al., 2009).

Our analysis used the entity linking approach introduced by Cornolti et al. (2013) to identify the evolution of topics in cloud computing technology. The approach of entity linking, in contrast to other methodologies, instead of using a bag of words concept, it tries to identify meaningful sequences (mentions) and can be annotated to specific identifiers (entities) retrieved from a catalog. We utilized Wikipedia as one of the most popular catalogs in entity linking in this study. We used TAGME software to exert the entity linking for Cloud Computing documents. It is a powerful tool that can be used to extract meaningful short phrases from an unstructured text and link them to Wikipedia entities (Ferragina and Scaiella, 2010). We called TagMe's API version 0.1.3 using Python programming language version 3.9.7 and implemented the process through a Jupyter notebook version 4.6. Following the study by Cuzzola et al. (2015), to cover many Cloud Computing technology relevant terms, we merged $n$ abstract and a title of each article as a text. The values for the area under the curve $F$-score were set as the stochastic setting of tunable parameters (Epsilon 0.427, q = 0.16 and Long_text 10). Because of the high volume of the documents being analyzed (49,828 articles), we used NVIDIA GPU Tesla P100, with 60GB memory. This process adds a column with the name "annotations" to the original dataset containing topics generated by the Entity Linking algorithm to each record. In the final step, after implementing TAGME, we removed topics that are not meaningful in the context of the study. For example, we removed Emerald, Elsevier, Hungary, and Budapest related to copyright information at the end of the articles.

To discover trends in topics, we used Mann-Kendall's test analysis after creating a topic using entity linking method. The Mann-Kendall test is used to identify hot topics within an academic field at a specific moment in time (as clarified later). The hot topics are those for which there are an increasing number of publications (Nederhof and van Wijk, 1997). The identification of hot topics can allow researchers to focus their attention on increasing areas of research instead of ones that are in decline (Marrone, 2020). Additionally, we used Kleinberg's burst detection algorithm to identify the popularity of topics over time. Kleinberg (2003) observes that hot topics initially appear, gain intensity over time, and then gradually dissipate. To find the hottest topics in cloud computing, we used the Mann-Kendall test as well as the burst detection algorithm as appropriate methods of inquiry (cf. Coccia, 2018b). It should be noted that the results of the two analyses can differ since Mann-Kendall's test tries to find topics that increase swiftly, whereas the Burst algorithm tries to show topics that are increased in terms of the frequency of occurrences in a text over a period (become hot) and then their importance gradually decreases.

- Mann-Kendall's test

One of the most widely used statistical tests for finding trends in the time series is Mann-Kendall (MK) test (Marrone, 2020; Zhang and Lu, 2009). In this non-parametric test, the correlation between the rank order of the observed values and their temporal order in time is considered (Hamed and Rao, 1998). The null hypothesis is that the sample data are independent and randomly distributed, or there is a serial correlation among the observation or data sample has not a monotonic trend (Hamed and Rao, 1998; Marrone, 2020; Zhang and Lu, 2009). We used pyMannKendall python package Version 1.4.2, which is a pure Python implementation of non-parametric Mann-Kendall trend analysis (Hussain, 2019). Studies show that non-parametric tests require that data are independent, and, in several situations, the observed data are likely autocorrelated, resulting in misinterpretation of trend-test results (Hamed and Rao, 1998) and an increased likelihood of falsely finding statistical significance without a trend being apparent, in the presence of positive serial correlation (Cox and Stuart, 1955). In this paper, we used the variance correction method for to consider serial autocorrelation as proposed by Hamed and Rao (1998). We used Theil-Sen estimator to estimate the magnitude of trends identified by Mann-Kendall test (Zhang and Lu, 2009). The absolute value of slope indicates the speed of change in trends. In this study, we used the slope to identify which topic trends are increasing rapidly.

- *Topic popularity using Burst Detection*

We used Kleinberg's burst detection algorithm to identify the bursty topics from the cloud computing publications. Kleinberg (2003) developed an algorithm for identifying topics that appear, boost intensively over time, and then gradually disappear. In other words, this algorithm identifies periods in which a term is uncharacteristically frequent or "bursty." According to Marrone (2020), we can use the Kleinberg algorithm to identify bursts in topics and detect hot (increasing) topics or cold (decreasing) topics over time (Marrone, 2020). We used SCI2 tool V1.3 to implement the Kleinberg burst detection algorithm and visualize the temporal bar graph on the cloud computing papers (Sci2 Team, 2009).

## 4. Results

### 4.1. Topic trends

Table 1 shows the top 20 most quickly increasing topics regarding their frequency of use in the Cloud Computing research, deploying the Mann-Kendall test. According to this technique, a distribution can be considered as a monotonic trend if the $p$-value is less than 0.05. Moreover, we calculated the $Z$ score as an indicator of normalized test statistics, and if the $Z$ is positive, there exists a strictly increasing trend. We also used the slope to find the top 20 quickly increasing trends.

**Table 1.** Most quickly increasing topics in cloud computing based on Mann-Kendall test results, 2004-2021 period

| Topics | $p$-value | Z-score | Slope | Topics | $p$-value | Z-score | Slope |
|---|---|---|---|---|---|---|---|
| Internet of Things | 0.00003 | 5.143 | 174.00 | Cryptography | 0.00004 | 5.060 | 29.22 |
| Mathematical optimization algorithm | 0.00001 | 5.336 | 136.42 | Parallel computing | 0.00002 | 3.699 | 26.53 |
| Virtual machine | 0.0001 | 3.887 | 93.66 | Machine learning | 0.00002 | 5.214 | 26.50 |
| Computer network | 0.00001 | 4.878 | 87.14 | Mobile cloud computing | 0.0005 | 3.450 | 23.45 |
| Encryption | 0.00004 | 4.600 | 82.90 | Fog computing | 0.000007 | 3.972 | 21.00 |
| Big data | 0.00005 | 4.063 | 75.84 | Artificial intelligence | 0.00009 | 5.333 | 20.40 |
| Distributed computing | 0.0001 | 4.040 | 53.50 | Computer architecture | 0.0001 | 3.811 | 20.00 |
| Load balancing (computing) | 0.0001 | 4.573 | 34.08 | Smart city | 0.00003 | 4.677 | 15.16 |
| Real-time computing | 0.0001 | 4.955 | 32.66 | Computer cluster | 0.00007 | 3.354 | 15.08 |
| Sensor | 0.0001 | 4.563 | 31.33 | Particle swarm optimization | 0.00001 | 4.953 | 14.38 |

**Note**: Z-score is a normalized test statistic of Mann-Kendall test; $p$-value is a significant level of test $p<0.5$. The items are with an in decreasing order according to the slope. A high slope indicates that the topic increases faster than topics with a small slope. The analysis provides a lot of topics but here are presented only the top twenty for clearness.

Table 1 shows that in the domain of cloud computing "Internet of things" with the highest slope value of 174.00 has been growing fast compared to the other topics. Interestingly, this result illustrates that among the relevant entities, Internet of things studies have the highest growth rate in multidisciplinary research of Cloud Computing. "Mathematical optimization" with the slope value of 136.42 has also the most significant growth rate in the number of publications in cloud computing. Accordingly, in Cloud Computing, Mathematical Optimization plays a vital role in developing the fundamental analytical aspects of this emerging technology. After that, "Virtual Machine" is the third quickest topic in terms of popularity in cloud computing with a slope value of 93.66. Since all the topics hold a positive significant level of $p$-value and $Z$ score, all trends are upward, but those topics with a slope value higher than mean (average of top 20 quickest topics slope values=50.54) can be considered as the hottest topic, which has gained momentum these years. In this regard, "Computer network" with a value of 87.143,

"Encryption" with a value of 82.91, "Big data" with 75.85, and "Distributed computing" with a slope value of 53.5 are the most fast-growth topics in the field of Cloud Computing.

Table 1 also shows that the "Particle swarm optimization" with 14.38, "Computer Cluster" with 15.08, and "Smart City" with 15.17 have the lowest slopes. This result suggests that although these topics are gradually growing at a stable rate, but not fast enough to consider them hot topics in Cloud Computing. These findings reveal a technological competition between Cloud systems infrastructure, hardware development side, and computing and software development to play the main role in technological evolution of Cloud Computing.

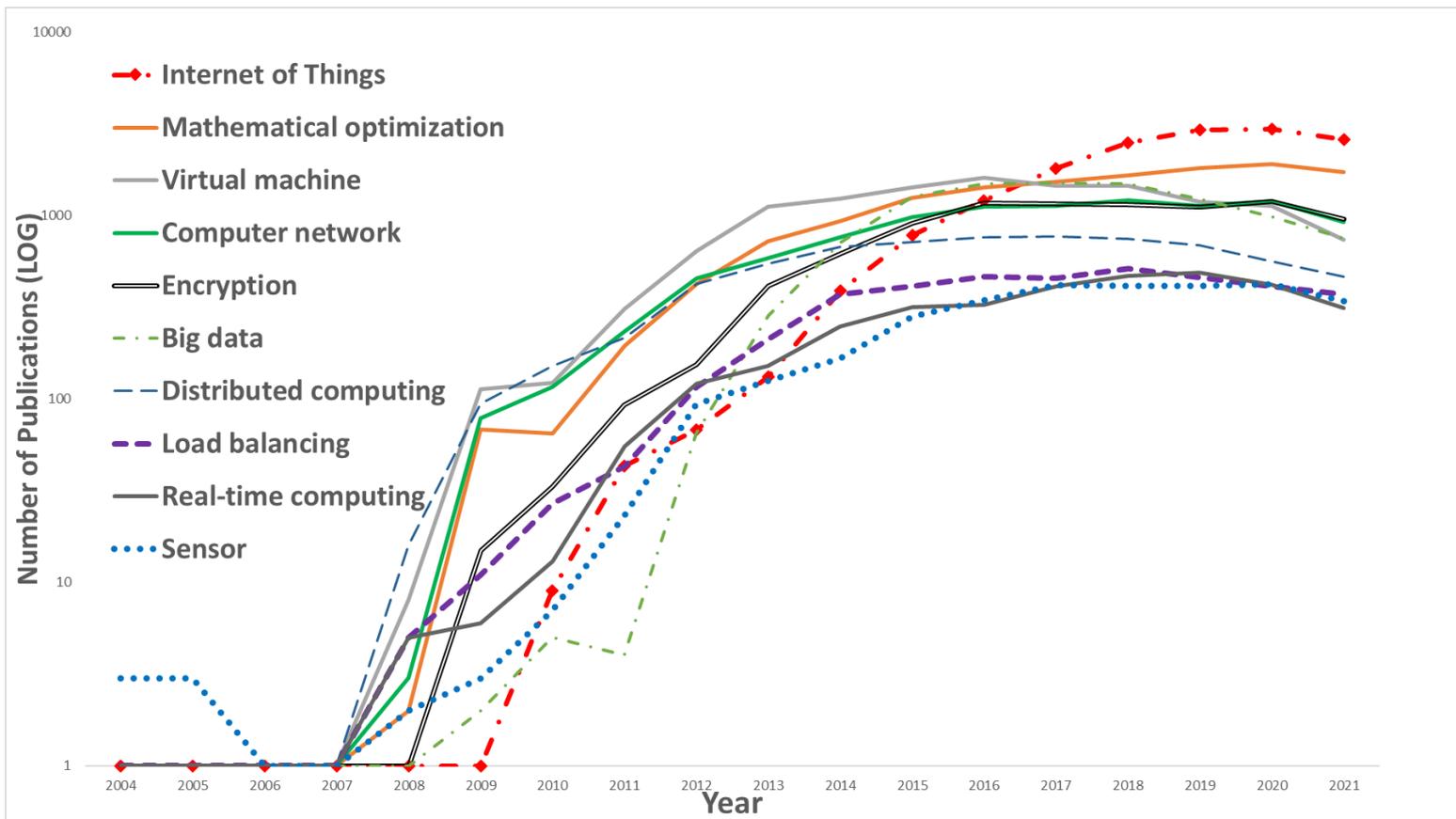

**Figure 1. Top 10 Cloud Computing related technologies based on the Mann-Kendall test**
*Note.* All items in the legend indicate topics. The analysis provides a lot of topics but here are presented only the top ten for clearness.

Figure 1 displays that "Distributed Computing" from 2008 had the highest number of publications compared with other topics in the field of cloud computing. After that, "mathematical optimization", "Virtual machine" and "Computer Network" initiated their popularity. We can consider these fields as leading topics until 2014 when Big data quickly accelerated, and until the middle of 2016, this topic gained significant interest for scientific research in Cloud Computing. It is also interesting to point out that the trend of "Internet of things" has the year 2012 as a main point of acceleration: the number of publications speeded up and increased faster than the other topics. The pathway of growth continued until 2021 when this field is the most popular in the research field of Cloud Computing . In addition, the topic of "sensor" is growing gradually, but still at a lower rate than other topics. Although this topic's popularity is straggled from other fields, its application as a complementary technology in critical topics is considerable and it can open new technological directions in cloud computing in process of technological interaction with Internet of things generating a fruitful co-evolution.

## 4.2. Topics Bursts

Figure 2 (A-B) visualizes the results of Kleinberg's bursting algorithm. This figure illustrates the top 100 topics in the Cloud Computing research over time. As seen in figure 2 (A-B), each topic has a start point (the year that burstness of the topic started) and an endpoint (the year that burstness of the topic ended). The distance between these two points indicates the length of topics' burstness. Moreover, the thickness of the bar of each topic in figure 2 indicates the weight of that topic (frequency of occurrences of the topic in the text) in all documents.

The first detected bursts in this study started in the year 2008. Based on the weights of topics, the most important topic in this year is Grid Computing with a length of 5 years from 2008 to 2012. Service Oriented Architecture is in the second rank regarding the weight of topics. All these topics roughly represent the evolution of infrastructure and computation foundations of cloud computing technology. In later period, from the year 2018, "fog computing", "artificial network", and "remote sensing" boosted, and from the computation side, complementary technologies started gaining momentum in the cloud computing research. The topics of "edge computing", "internet of things", "deep learning", "machine learning", "blockchain" and "reinforcement learning" emerge from 2019 to 2020 and are the most weighted topics according to the analysis presented in figure 2, which ultimately shows the evolution of knowledge behind the cloud computing from mathematics and infrastructure to computing and interdisciplinary. finally, the topics of "quantum computing", "particle swarm optimization", "elliptic-curve cryptography" and "ciphertext indistinguishability" burst in 2021 and can be considered as the most emerging topics in the research field of cloud computing.

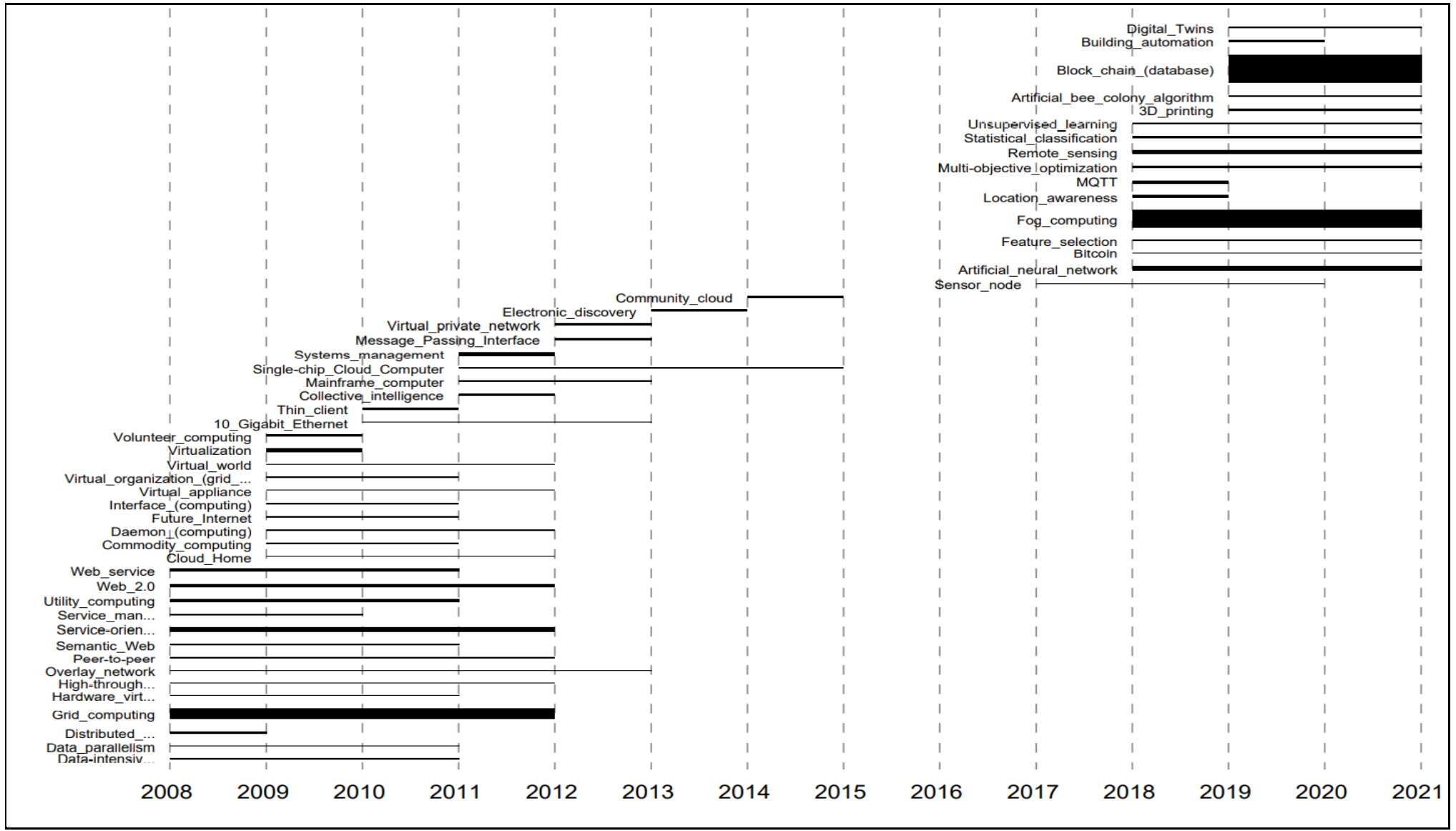

**Figure 2A.** Bursty topics in Cloud Computing based on Kleinberg's burst detection algorithm. *Note.* The *x*-axis indicates years from 2008 to 2021. Each bar has the start year and end year that emerge and disappear over time. The thickness of bars indicates the weight of topics according to their frequency.

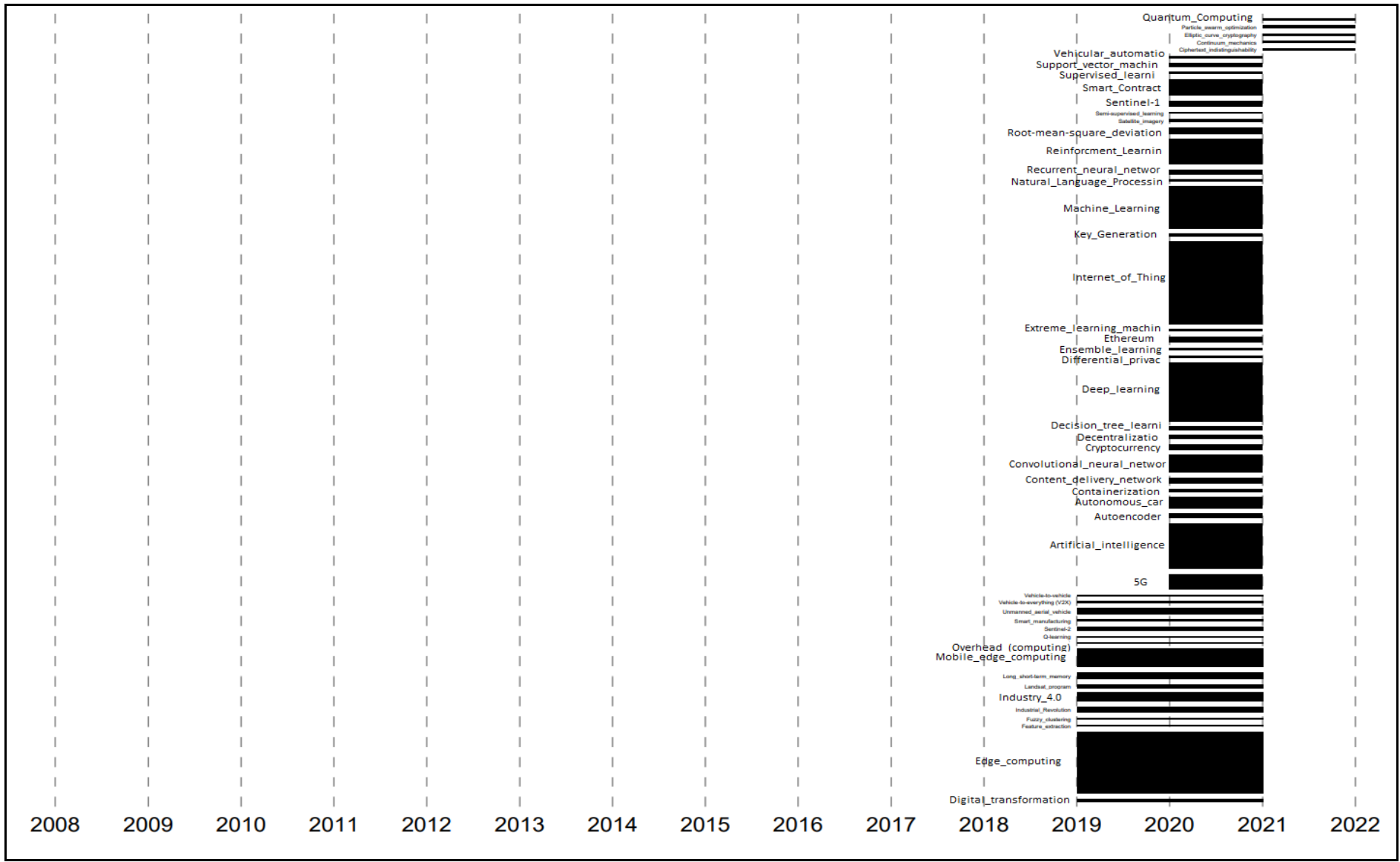

**Figure 2B.** Bursty topics in Cloud Computing based on Kleinberg's burst detection algorithm. *Note*. The x-axis indicates years from 2008 to 2022. Each bar has the start year and end year that emerge and disappear over time. The thickness of bars indicates the weight of topics according to their frequency.

## 5. Discussions

Results of the study show that technological ecosystem of cloud computing as a General-purpose technology is mainly driven by research fields of "internet of things", "mathematical optimization" and "virtual machine". Other topics playing a main role are "computer network", "encryption", "big data", and "distributed computing". In particular, "distributed computing" from 2008 has high level of publications, whereas from 2014 Big data quickly accelerated associated to the evolution of IT technology, e.g., software development, that has generated interaction with different technologies (e.g., powerful computer, Internet of Things, etc.) and co-evolutionary pathways of growth (in fact, data can be collected from mobile applications, social networks, websites, etc.); finally the trend of "Internet of Things" from the year of 2012 has a remarkable acceleration. One of explanations of high acceleration of the Internet of Things is because this new technology produced so many data and cloud computing algorithms supports the computation and analysis of this big data. A similar explanation is also for virtual machine (cf., Table 1 and Figure 1). In general, technological trajectories of cloud computing have a takeoff from 2007 to 2009, whereas since 2019 it seems that they are in a phase of technological maturity. Results also show the main evolution of the infrastructure and computation foundations of cloud computing technology. Moreover, results show that that from the year 2018, "fog computing", "artificial network", and "remote sensing" boosted, instead the topics of "edge computing", "internet of things", "deep learning", "machine learning", "blockchain" and "reinforcement learning" emerge from 2019 to 2020 suggesting that the evolution of knowledge in cloud computing is more and more based on mathematic and software development, and new infrastructures and equipment. Finally, research topics "quantum computing", "particle swarm optimization", "elliptic-curve cryptography" and "ciphertext indistinguishability" burst in 2021 and can be considered as the current emerging research fields in cloud computing technologies.

☐ *Theoretical implications.*

Technological trajectory is the activity of technological process along the economic and technological trade-offs of the general-purpose technology of cloud computing is driven by technological paradigm of Information and Communication Technology, ICT (Dosi 1988; Nelson and Winter 1982). A major impulse to innovation in cloud computing that characterizes new technological trajectories (or avenues) just described, it is due to the scientific activity in these topics that becomes "endogenized" to support a scientific and technological accumulation and a search of profit-motivated for businesses (Dosi, 1988). Moreover, evolution of cloud computing technologies can be due to interaction with new processes of innovation in software development, computer technology, Internet of Things, etc.

Technological trajectories in cloud computing are driven by technological paradigm of Information and Communication Technology, ICT (Dosi 1988; Nelson and Winter 1982). A major impulse to innovation in cloud computing that characterizes new technological trajectories (or avenues) just described, it is due to the scientific activity in these topics that becomes "endogenized" to support a scientific and technological accumulation and a search of profit-motivated for businesses (Dosi, 1988). Moreover, evolution of cloud computing technologies can be due to interaction with new processes of innovation in software development, computer technology, Internet of Things, etc.

Main characteristics of technological analysis here that can support the development of the general-purpose technology of cloud computing are:

- *Specialization* systems, such as in healthcare and encryption based on the interaction of hardware and software elements and big data needed to enable cloud computing.
- *Development of specific infrastructures*, such as computing power, servers, network switches, memory and storage clusters, as well as new interfaces for users to access to virtual resources.
- Cloud systems foster the *integration of complementary elements* of hardware and software (e.g., Internet of Things, virtual machine, etc.) for a common platform and technological ecosystem that can manage multiple clouds fostering the co-evolution with a technological parasitism and symbiosis (Coccia, 2019, 2019a; Coccia and Watts, 2020).
- Exploitation of *economies of scale* because cloud infrastructure supplies the same capabilities as physical infrastructure but with additional advantages given by a lower cost, greater flexibility, and scalability.
- *Economies of standardization* creating similar software, interfaces, platforms and infrastructure for extensible cloud systems, fostering interaction with different technologies and adopters (Choi et al., 2011; Katz and Shapiro, 1985; Shapiro and Varian, 1999).

☐ *Principal innovation policy and innovation management implications*

The description of characteristics of evolutionary pathways in the general-purpose technology of cloud computing here can improve the allocation of R&D investments in private and public organizations for beneficial social impact (Coccia and Rolfo, 2000; Roshani et al., 2021; Pagliaro and Coccia, 2021). In fact, the evolution of cloud computing technologies can have numerous applications in emerging markets. Policymakers and scholars know that financial resources can be an accelerator factor of progress and diffusion of science and technology to support the scientific and technological development (Roshani et al., 2021). This study shows critical research fields and technologies in cloud computing that are growing; policy makers can allocate economic resources towards these research fields having a high potential of growth to support the evolution of cloud computing technology in interaction with other technologies (e.g., Internet of Things, virtual machine, encryption, big data, new sensors) generating co-evolutionary for positive impact in science and society. In fact, these findings can support policymakers and funding agencies in making efficient decisions regarding sponsoring specific research fields and technological trajectories in cloud computing that can accelerate the development of science and technology and foster technology transfer having fruitful effects in markets for wellbeing of people in society.

6. **Conclusions**

The evolution of cloud computing technology is basic to explain main drivers of technological and social change in markets (Cresswell et al., 2022). The study here reveals that cloud computing is mainly driven by interacting with main research fields of "internet of things", "virtual machine", "computer network", "encryption" and "big data. Latifian (2022) argue that the development of new businesses by the utilization of big data can be effectively addressed using cloud computing and related computing technology. . In fact, cloud computing is a powerful technology for storing, compressing, analyzing and processing big data. However, these conclusions are, of course, tentative. The rapid evolution of cloud computing technologies has a lot of potential commercial applications, but it can also raise some problems during the implementation in organizations and society. One of the problems in the evolution of cloud computing is barriers to progress given by transaction cost of data migration, lack of know-how for a vast implementation in provider organizations, a lack of standardization and central regulations (Swann, 2000; Zhang et al., 2020). Other problems during the implementations of cloud computing technology may be associated with aspects of ethics, privacy and civil rights (cf., Batra et al., 2021; Holter et al., 2021; Inglesant et al., 2021 for similar problems in quantum computing).

Overall, then, although this study has provided some interesting, but preliminary results, it has also some limitations. First, the search queries are affected by ambiguous meanings of terms in the cloud computing domain. Second limitation is that publication analysis can only capture certain aspects of the evolution of cloud computing d technology. Third, there are confounding factors that affect the dynamics of cloud computing, such as equipment of IT platform, skilled human capital, available big data, etc. Despite these limitations, results here show main technological trajectories in cloud computing that provide new knowledge of the evolution of this new technology for supporting R&D investments in research fields having a high potential of growth in society, such as internet of things, virtual machine, big data management, encryption, etc. (Roshani et al., 2021). To conclude, results presented here clearly point out the need for more detailed technology analyses to clarify the structure and evolution of cloud computing technology to support further implications for innovation management and research policy that accelerate the development of new science and technology with fruitful effects for future wellbeing of people in society.